# The Impacts of Learning Assistants on Student Learning of Physics

Jada-Simone S. White[1], Ben Van Dusen[1], & Edward A. Roualdes[2]
[1]*California State University Chico, Department of Science Education*
*101 Holt Hall, Chico, CA, 95929, USA*
[2]*California State University Chico, Department of Mathematics and Statistics*
*204 Holt Hall, Chico, CA, 95929, USA*

**Abstract.** *This study investigated whether and how Learning Assistant (LA) support is linked to student outcomes in Physics courses nationwide. Paired student concept inventory scores were collected over three semesters from 3,753 students, representing 69 courses, and 40 instructors, from 17 LA Alliance member institutions. Each participating student completed an online concept inventory at the beginning (pre) and end (post) of each term. The physics concept inventories tested included the Force Concept Inventory (FCI), Conceptual Survey of Electricity and Magnetism (CSEM), Force and Motion Concept Evaluation (FMCE) and the Brief Electricity and Magnetism Assessment (BEMA). Across instruments, Cohen's d effect sizes were 1.4 times higher, on average, for courses supported by LAs compared to courses without LA support. Preliminary findings indicate that physics students' outcomes may be most effective when LA support is utilized in laboratory settings (1.9 times higher than no LA support) in comparison to lecture (1.4 times higher), recitations (1.5 times higher), or unknown uses (1.3 times higher). Additional research will inform LA-implementation best practices across disciplines.*

## I. INTRODUCTION

A central goal of the LA model is to improve undergraduate STEM student learning outcomes by increasing faculty use of research-based instructional strategies in undergraduate courses [2]. Despite the models growth and apparent success, there are a limited number of studies that evaluate the impact of LAs on student learning. With few exceptions [3], the existing literature on the LA model only examines the impact of LAs at individual institutions and typically with individual instructors. The findings of these studies show promising results in specific settings but do not examine the systemic impact of LAs across institutional and classroom contexts.

The LA Alliance is an international network of institutions that have established, or are interested in starting, LA programs. The Alliance was created to support the dissemination, sustaining, and scaling of LA programs nationally and internationally. The Alliance is currently composed of over 90 institutions, each with their own set of institutional contexts that shape the way that LAs are implemented within courses. For example, courses utilize LAs in vastly differing formats, ranging from supporting required laboratory and lecture portions of the course, to more informal recitations, workshops, and tutorials outside of lecture. The intrinsic variation in institutional and classroom contexts can act as a barrier to abstracting study results and reliably scaling course transformations across settings. The creation of the LA Alliance has made it feasible to examine student outcomes across many institutional contexts. The LA Supported Student Outcomes (LASSO) online assessment tool was created specifically to facilitate the large-scale examination of student learning across classroom contexts (see methods section for details). An increasing number of institutions without LA programs are also utilizing LASSO, as well as other DBER resources available online through the LA Alliance, allowing comparisons of courses with and without LA support nationwide. Previously, we analyzed the LASSO dataset to document the broad trends in student outcomes in LA supported courses across disciplines [3]. In this paper we examine whether student outcomes vary depending on how LAs are implemented specifically in physics courses.

## II. RESEARCH QUESTIONS

By examining student outcomes and physics classroom features across institutional contexts we investigated the questions: (1) How does the presence of LAs impact student performance in physics courses, if at all? (2) How do specific uses of LAs impact student performance in physics courses, if at all?

## III. LITERATURE REVIEW

Many investigations into student learning in LA-supported courses have utilized concept inventories, such as the Force Concept Inventory [4], to measure students' disciplinary knowledge in the first week of a class (pre-test) and the last week of the class (post-test). Paired scores are then typically used to calculate either the normalized learning gain or Cohen's d effect size. The normalized learning gain is a measure of student improvement (post%-pre%) divided by the amount they could improve (1-pre%) [5]. Cohen's d is a measure of improvement (in this case from pre to post scores) in units of standard deviations at the course level [6].

Using these analytical techniques, researchers have associated LAs with improved student learning in university science and math courses. The introduction of LAs was associated with significantly improved student outcomes in chemistry courses with a research-based curriculum [7]. In physics, the use of LAs to support research-based curriculums, such as Tutorials [8], has been associated with improved student learning [9]. Students in the LA-supported physics course were also shown to have improved outcomes in subsequent physics courses [10]. In a calculus class with LA-supported oral assessments, the gap between students who had been labeled "at risk" due to placement scores and their peers was eliminated by the end of class [11].

These studies are very encouraging and have driven much of the growth of the LA model. While each of these investigations indicate that LAs are making an impact in the courses included in the study, it can difficult to identify the specific class features that had the most impact and should be scaled across institutions. The LASSO project was launched to identify large-scale and discipline specific trends in student outcomes that could be used to empirically develop a set of best practices for LA-implementation.

In the initial examination of LASSO data, hierarchical linear models nested student level data (e.g. gender, race, and weekly time spent with LAs) within course level data (e.g. student/LA ratio and discipline) across biology, chemistry, and physics courses: Cohen's d effect size was used as an estimate of learning outcomes under various educational contexts [3]. For LA-supported courses, the mean effect sizes of students who spent 16-30 min/week interacting with LAs were more than twice as large as their peers who spent 0 min/week interacting with LAs [3]. Interestingly, mean effect sizes improved with the number of times that an instructor had previously taught the course using LAs [3].

This publication builds on the analysis of the growing LASSO dataset to identify trends in student outcomes associated with particular instantiations of the LA model in physics classes. Specifically, we evaluated the primary way in which LA support is implemented within physics courses as indicated by faculty (i.e., laboratory, mandatory or optional recitations outside of lecture, or small groups < 50% or > 50% during lecture).

## IV. METHODS

### A. Data Collection

Physics concept inventory data were collected using the LA Supported Student Outcomes (LASSO) online assessment tool. LASSO is a free tool, hosted on the LA Alliance website [12], that allows all STEM faculty (LA-using or not) to easily administer Research-Based Assessment Instruments as pre- and post-tests to their students online. When using LASSO, faculty provided course-level information, selected their assessment(s), and uploaded a list of student names and emails. After faculty launched an assessment, their students received emails with unique links to complete their pre-tests online. The LASSO system also allowed faculty to track their students' participation and send reminder emails. At the end of the semester students received another set of emails with unique links to their post-tests. Once completed, faculty had the opportunity to download their individual student's responses, as well as a summary report that showed the distribution of their students' pre and post scores, normalized learning gains, and effect size (Cohen's d). As of the Fall 2016 semester, LASSO is hosting 15 online instruments across the STEM disciplines.

### B. Data Analysis

In this investigation we examined data from physics courses that used the Force Concept Inventory (FCI) [4], Conceptual Survey of Electricity and Magnetism (CSEM) [13], Force and Motion Conceptual Evaluation (FMCE) [14], and Brief Electricity and Magnetism Assessment (BEMA) [15]. Over the first three semesters of data collection, LASSO collected > 8,500 unique student pre- and post-test responses on the four instruments from 143 physics courses at 19 institutions nationwide. Data were cleaned in a four-step process. First, all student responses with answers to less than 80% of the concept inventory questions were removed. Second, any student responses that were not part of a matching pre-post set were removed. Third, any classes that were left with less than 10 matched sets of student responses (either due to low enrollment or participation) were removed. Finally, all unrealistic effect sizes (≤ -1.0 or ≥ 4.0) were removed. Once student results were cleaned, there were 3,740 usable pre-post pairs of responses from 69 courses at 17 institutions (Table I). Each response was scored and the course-level effect size (Cohen's d) was calculated for each student. Cohen's d is a measure of change (in this case from pre to post scores) in units of standard deviations at the course level.

**TABLE I.** Cleaned data counts.

| Concept Inventory | Institutions | Courses | Students ($N_{Paired}$) |
|---|---|---|---|
| FCI | 9 | 26 | 697 |
| FMCE | 9 | 15 | 1,592 |
| BEMA | 4 | 7 | 680 |
| CSEM | 4 | 21 | 754 |
| **Total** | **17** | **69** | **3,753** |

To answer the first research question, we tested the difference in course mean (Cohen's d) effect sizes in the absence (N=18) and presence (N=51) of LA support using a Welch two-sample t-test. To answer the second research question, we used course-level information provided by instructors indicating the primary activity that LAs facilitated within the courses. To compensate for uneven sampling, prior to analyses we binned mandatory (N=8) and optional (N=1) recitations; as well as the use of small-groups for < 50% (N=2) and > 50% (N=3) of the time in lecture, into single categories (i.e., recitation and lecture, respectively). Using a simple ANOVA, we evaluated effect size as a function of five categories of LA implementation: None (N=18), Laboratory (N=4), Recitation (N=9), Lecture (N=5), or Unknown (unspecified; N=33). In addition to checking normality and homoscedasticity visually, we used Levene's Test to verify the assumption of homogeneity of variance in the presence of uneven sampling among categories. For post-hoc multiple comparisons, we used t-tests with Bonferroni correction and verified results with Tukey's HSD. Figures report the 95% Confidence Interval (±1.96*S.E.) to aid visualization of significant results. All analyses were conducted using R 3.0.2 GUI 1.62 (©2012, R Foundation for Statistical Computing) using the following packages: base, car, gdata, and gplot.

## V. FINDINGS

The presence of LAs was associated with improved student outcomes (Figs. 1 & 2). Figure 1 illustrates that courses without LAs were found to have mean effect sizes significantly lower than those of courses with LAs. On average, the effect size of courses supported by LAs was 1.4 times higher than the effect sizes of courses lacking LA support (Fig. 1: t-test, $t_{28.03}$=-2.7125, p=0.01; Levene's Test: $F_{1,67}$=7e$^{-4}$, p=0.98). To evaluate the impact of specific LA-supported activities, we used the R default, treatment contrasts, to set our baseline category of No LAs (Intercept = None) and compared with the mean differences of each specific LA-uses. Overall, the average course effect size varied significantly with the primary LA-supported activities, relative to courses without LA support (Table II).

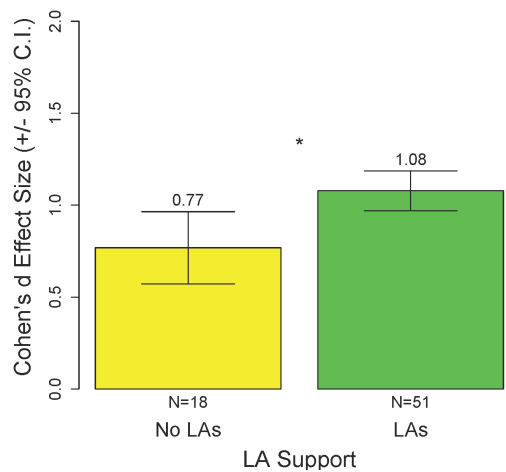

FIG. 1. Mean Effect Size as a function of LA Support.

In support of graphical depictions (Figs. 1 & 2), courses with no LAs were estimated to have an effect size of 0.77 (Table II). All categories that included LAs increased positively (ranging from 0.25 to 0.67 higher effect sizes) relative to the baseline (Intercept). Note that there was a larger estimated increase in the Lecture, relative to the Unknown category, however there was too much variation within the limited N to register as different from courses with no LAs. Post-hoc comparisons indicated that were no significant differences in the effect size among the primary uses of LA support tested (Fig. 2: green bars, not including category None; Tukey HSD and t-tests with Bonferroni correction).

TABLE II. Effect size as a function of primary use of LAs support in physics classes, in comparison to no LAs (Intercept = None).

| | Coefficient Estimate | S.E. | t value | p-value |
|---|---|---|---|---|
| (Intercept) | 0.77 | 0.09 | 8.203 | <0.001*** |
| **Laboratory** | 0.67 | 0.21 | 3.046 | <0.01** |
| **Recitations** | 0.38 | 0.16 | 2.345 | <0.05* |
| Lecture | 0.32 | 0.20 | 1.613 | 0.11 |
| **Unknown** | 0.25 | 0.11 | 2.101 | <0.05* |

ANOVA: $F_{4,64}$ = 3.132, p=0.02; Levene's: $F_{4,64}$ = 0.7927, p=0.53.

The other primary uses of LA support within physics courses obtained generally higher average effect sizes, relative to courses with no LAs (Table II): Recitation (1.5 times higher), Lecture (1.4 times higher) and Unknown (1.3 times higher)(Fig. 2). However, the learning outcome of students in physics courses who utilized LA support in the laboratory was a significantly higher, at nearly twice (1.9 times) the mean effect size of courses with no LA support (Fig. 2). Follow-up comparisons using Tukey's HSD and Bonferroni correction for multiple comparisons (p<0.0125) indicated this was the only statistically significant difference when robust methods were applied (t-test, $t_{8.93}$=-4.502, p=0.0015; Fig. 2: *p<0.05; Bonferroni).

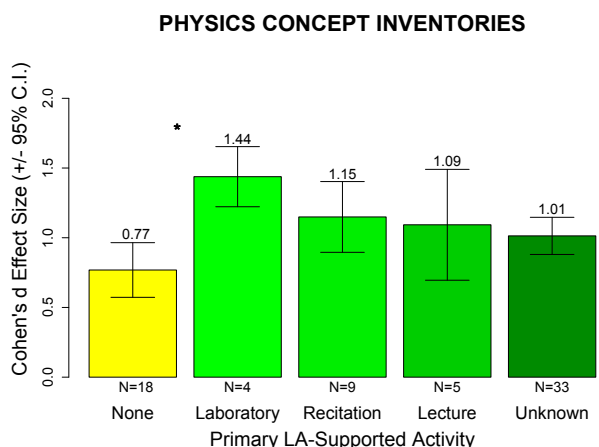

FIG. 2. Effect size as a function of primary use of LAs.

Thus, given the variable sample sizes among primary LA use categories, we did not have adequate statistical power to state differences were significant when robust methods were applied.

## VI. DISCUSSION

Evidence is mounting that using LA-supported activities to teach core physics concepts is more effective than traditional approaches, which lack LAs. Our study is the first to compare physics courses from across the nation. We utilized robust statistical methods to test whether physics courses with LA support are more effective than courses with no LAs overall, and across various implementations of LA-support (i.e., Laboratory, Recitation, Lecture, or Unknown). By dividing the mean of the category of interest by the mean of our baseline category (no LA or None), we find that the mean Effect Size of courses with LAs was 1.3 – 1.9 times higher than courses without LAs (Fig. 2). Using LAs to support activities within physics labs appears to be the most beneficial, followed by Recitations. Interestingly, lecture showed the most variation relative to the mean (Table II), suggesting there is more variation in how LAs are implemented in lectures. The large number of courses in the "Unknown" LA-use category was unfortunate and we are working with LASSO Technicians to improve faculty completion rates.

## VII. CONCLUSION & FUTURE WORK

Our evaluation of physics courses from 17 institutions nationwide indicated that LA-support is advantageous in promoting understanding of core physics concepts, as measured by the FCI, FMCE, BEMA, and CSEM. The mean Effect Size ranged from 1.3-1.9 higher in courses supported by LAs, relative to courses without LAs (Fig. 2). There were no significant differences among the LA-use categories tested, reinforcing that all implementations are beneficial, possibly because they promote equity in the physics classroom [16]. Note that these analyses include only three semesters of data collected from LASSO. As more data are collected, the statistical power to identify specific activities that best promote understanding of core concepts within and among physics instruments, as well as other disciplines, will also grow. Thus, to promote the advancement of PER, and DBER in general, you are invited to use the LASSO online tool [12].

This paper is contribution No. LAA-004 of the International Learning Assistant Alliance. We gratefully acknowledge that the NSF-DEW #1525338 funded this LASSO research.

[1] C. Wieman, speech, (Oct 1st, 2014).
[2] V. Otero, American Physics Society, (2016).
[3] B. Van Dusen, et al., PERC 2015 Proc. (2016).
[4] D. Hestenes, M. Wells, & G. Swackhamer, Force concept inventory. *The Physics Teacher*, *30*(March), 141–158 (1992)
[5] R. Hake. *American Journal of Physics*, *66*(1), (1998)
[6] J. Cohen, *Statistical power analysis for the behavioral science.* Academic Press. (2013).
[7] L. Langdon, *Presented at the 247th ACS National Meeting*, Dallas, TX, March (2014).
[8] L. McDermott and P. Shaffer, *Tutorials in introductory physics*. Prentice Hall. (1998).
[9] S. Pollock and N. Finkelstein, Phys. Rev. ST Phys. Educ. Res., **4**, (2008).
[10] S. Pollock. *Physical Review Special Topics - Physics Education Research*, (2009).
[11] M. Nelson, PRIMUS: Problems, Resources, and Issues in Mathematics Undergraduate Studies 21, 1 (2011).
[12] www.learningassistantalliance.org retrieved 6/29/2015.
[13] Maloney, O'Kuma, etc, 2001
[14] R. Thornton, and D. Sokoloff, American Journal of Physics, **66,** 4 (1998)
[15] L. Ding, R. Chabay, B. Sherwood, and R. Beichner, Phys. Rev. ST Phys. Educ. Res, **2**, 1 (2006).
[16] B. Van Dusen, J.S.S. White, E. Roualdes. PERC 2016.